\documentclass[useAMS,usenatbib]{mn2e}

\usepackage{amssymb}
\usepackage[fleqn]{amsmath}
\usepackage{color}
\usepackage[bookmarks,bookmarksopen,colorlinks,linkcolor={red},citecolor={blue},urlcolor={green}]{hyperref}
\usepackage{graphicx}
\usepackage{natbib}
\usepackage{times}

\title[The Role of Plasma Instabilities in the Propagation of Gamma-Rays from Distant Blazars]{The Role of Plasma Instabilities in the Propagation of Gamma-Rays from Distant Blazars}

\author[A. Saveliev, C. Evoli and G. Sigl]
{A. Saveliev\thanks{E-mail: andrey.saveliev@desy.de}, C. Evoli\thanks{E-mail: carmelo.evoli@desy.de} and G. Sigl\thanks{E-mail: guenter.sigl@desy.de}\\
{II}. Institut f\"ur Theoretische Physik, Universit\"at Hamburg, Luruper Chaussee 149, 22761 Hamburg, Germany}

\begin{document}
\maketitle
\label{firstpage}

\begin{abstract}
The observation in the GeV band of distant blazars has been recently used to put constraints on the Extragalactic Background Light (EBL) and Extragalactic 
Magnetic Fields (EGMF). To support such claims one has to assume that the leptonic component of the electromagnetic cascade initiated by blazar gamma-rays is 
deflected away by strong enough EGMF, suppressing the signal in the Fermi window. Apart from magnetic fields, the development of such a cascade might be 
affected by plasma instabilities due to interactions with the ionized component of the Intergalactic Medium (IGM). In this paper we model the electromagnetic 
cascade through a Monte Carlo simulation in which both effects are taken into account separately, and we derive constraints on these scenarios from the 
combined Fermi-HESS data set. In the specific case of 1ES 0229+200 observations, we show that both explanations of the GeV flux suppression are compatible with 
the available data, specifically by assuming a magnetic field of $B \gtrsim 10^{-16}\,\rm{G}$ or an IGM temperature of $T \lesssim 5 \times 10^{4}\,\rm{K}$ 
along the line of sight. Future observations of the spectra of high redshift ($z\lesssim 1$) TeV objects will help to distinguish magnetic field and plasma 
effects on electromagnetic cascades in the IGM.
\end{abstract}


\begin{figure*}
\centering
\includegraphics[scale=0.58]{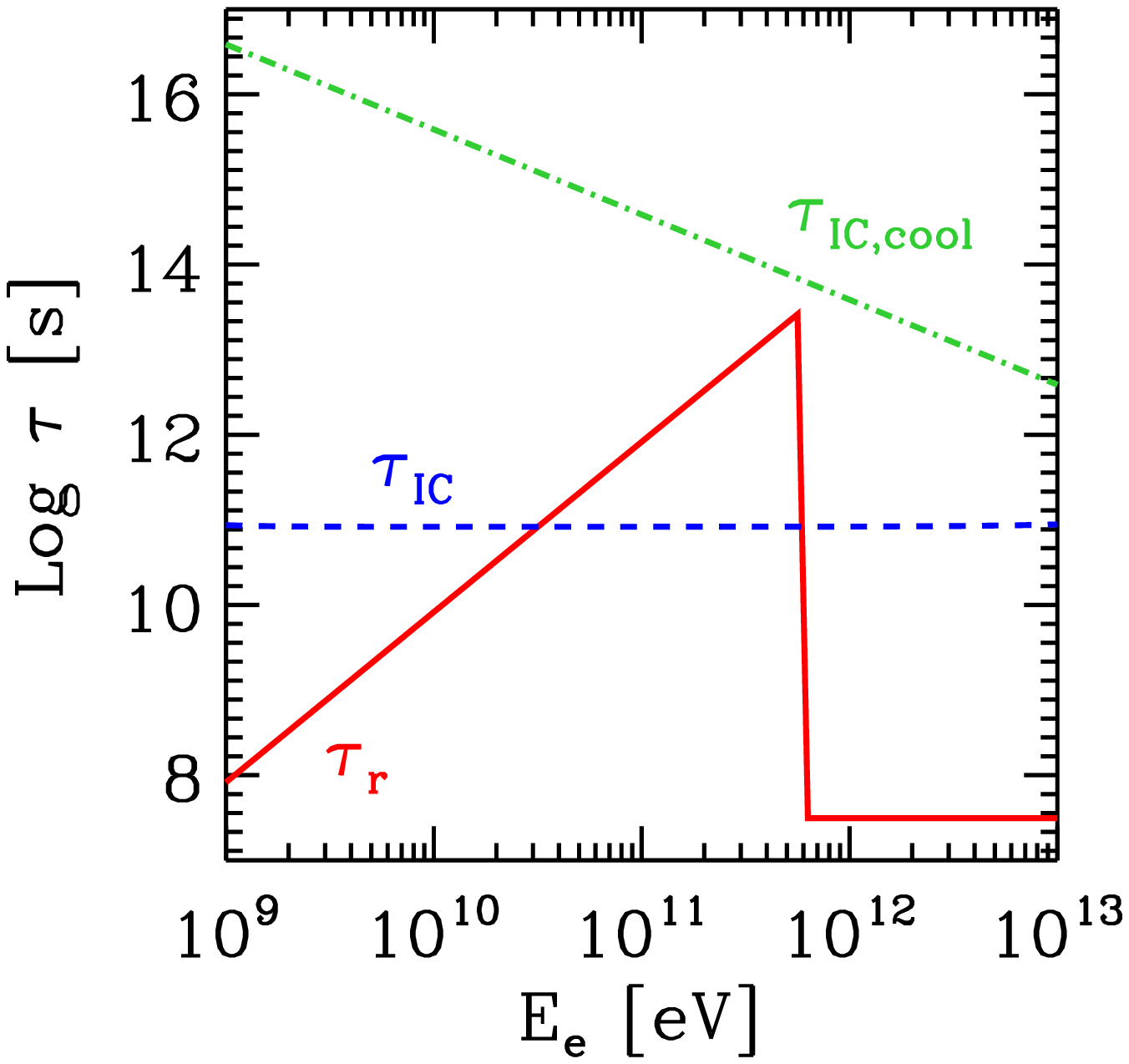}
\includegraphics[scale=0.58]{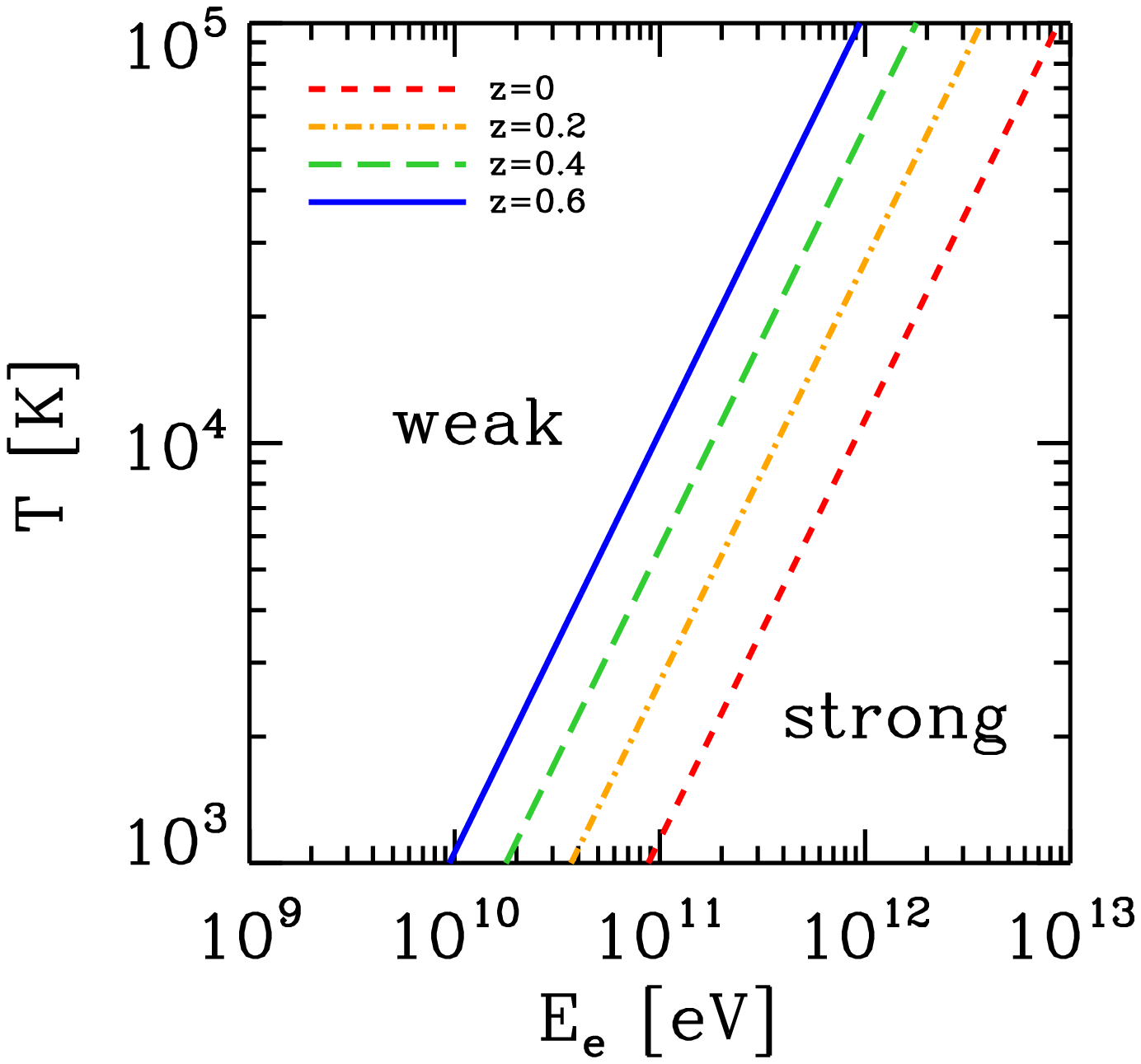}
\caption{
  \emph{Left panel}: The interaction time $\tau_{{\rm IC}}$ (dashed, blue) and the cooling time $\tau_{{\rm IC,cool}}$ (dot-dashed, green) for Inverse Compton 
  scattering and the relaxation time $\tau_{{\rm r}}$ for electrons due to plasma effects (solid, red). The parameters used here are 
  $n_{\rm{IGM}}=10^{-7}\,{\rm cm^{-3}}$, $T=10^{4}\,{\rm K}$, $z=0.14$ and $n_{\rm{beam}}$ according to Eq.~\ref{nBrod} with $L=10^{37.5}\,{\rm W}$. Note the 
  transition for $\tau_{{\rm r}}$ between the weak and the strong blazar regime at $E_{e} \simeq 5 \times 10^{11}\,\rm{eV}$. \emph{Right panel}: Transition 
  between the weak and the strong blazar regime (denoted by 'weak' and 'strong', respectively) given by the condition in Eq.~\ref{ncrit} for different 
  redshifts with $n_{{\rm IGM}}=10^{-7}\,{\rm cm^{-3}}$ and the beam density as for the left panel. For parameters forming the region below a given line the 
  system is in the strong  blazar regime with the modulation instability dominating, the region above the line gives the weak regime where NLD is most 
  important. The redshifts are $z=0$ (dashed, red), $z=0.2$ (dot-dashed, orange), $z=0.4$ (long-dashed, green) and $z=0.6$ (solid, blue).
}\label{fig:timescales}
\end{figure*}

\section{Introduction}
Up to the present day little is known about the origin, evolution and even the existence of Extragalactic Magnetic Fields (EGMF) making them a broadly 
discussed topic \citep[for a review see, e.g.,][]{PhysRep.348.163,DuNe}. 

On the other hand, theoretical predictions based on numerical simulations of structure formation predict that $\mathcal{O}({\rm nG})$ magnetic fields should be 
present in large scale structure filaments \citep[see, e.g.,][]{Ryu16052008} and, at the same time, tiny magnetic fields inside intergalactic void regions have 
been proposed as possible remnants of primordial magnetic fields that were generated in the early Universe \citep[for recent results see, for example,][]{Kahniashvili:2012vt,Saveliev:2013uva}. 
For these reasons the possible detection of EGMF is an important tool for our understanding of their origin and role in both the cosmology and astrophysics of 
structure formation.

Recently, various authors claimed to have found evidence for non-vanishing EGMF by analyzing the gamma-ray spectra of distant blazars 
\citep{2010MNRAS.406L..70T,Neronov02042010,Dolag:2010ni,2041-8205-733-2-L21,Taylor:2011bn,Vovk:2011aa} based on the methods developed in 
\citet{d'Avezac:2007sg,PhysRevD.80.123012}.
The common scenario is that gamma-rays emitted by a TeV source develop an electromagnetic cascade of which the charged component is deflected away from the 
line of sight by EGMF resulting in a suppression of the observable signal in the GeV range. Such suppression has been observed by combining HESS (as well as 
other Imaging Atmospheric Cherenkov Telescopes) and the Fermi Large Area Telescope (LAT) data of different objects 
\citep[see, e.g.,][]{1538-4357-696-2-L150,ReyesICRC2009,Acciari:2010qw,Taylor:2011bn}. 

Electromagnetic cascades develop due to the interaction of photons and electrons/positrons with the Extragalactic Background Light (EBL) of which in this 
context the most important components are the Cosmic Microwave Background (CMB) and the Infrared Background (IB). For photons the dominating process is the 
production of electron/positron pairs where, for energies up to approximately $10^{14}$~eV the reaction with the IB is the most important, making the interaction 
time scale highly sensitive to the given IB model. On the other hand, for the propagation of electrons or positrons of the cascade the dominating reaction is 
the Inverse Compton (IC) scattering on CMB. Here the charged particle upscatters a background photon to higher energies, therefore effectively injecting an 
additional photon into the cascade. 

By comparing the theoretical predictions of spectra for which this cascading process is influenced by magnetic fields to measurements of experiments like Fermi 
LAT the authors of the works mentioned above were able to derive lower limits on EGMF of the order of $B_{{\rm EGMF}} \ge 10^{-18} - 10^{-17}\,{\rm G}$. These 
results are still controversial to some extent as there are also claims that the observations are still in agreement with a vanishing EGMF hypothesis after an 
improved statistical analysis based on the combination of observations for different objects \citep{Arlen:2013vma}.

It should be noted here that complementary constraints on EGMF have been also derived by other methods than the described specific analysis of the spectra, 
e.g.~by considering the time-delayed echo of secondary gamma-rays produced in the cascade \citep{Plaga1994,1538-4357-686-2-L67}, by analyzing the extended 
emission of point-like sources \citep{JETPLett.85.10.473,0004-637X-703-1-1078,PhysRevD.80.023010,2041-8205-719-2-L130} or by assuming a model of secondary 
gamma-rays produced by cosmic ray interactions \citep{Essey:2010nd,PhysRevD.87.063002}.
 
An important aspect common to all the studies listed above is that they only take into account interactions with the photon backgrounds while ignoring the 
effects caused by the intergalactic medium (IGM). The latter, however, might be important as has been argued by \citet{0004-637X-752-1-22,0004-637X-758-2-102}: 
In fact, plasma instabilities may dissipate away large amounts of the cascade energy and effectively suppress the cascade development partially or completely, 
becoming the dominant energy loss mechanism. These plasma instabilities are created due to the fact that the mutual interaction of two fluid streams (the 
electron/positron beam generated in the TeV cascade and, on the other hand, the ionized IGM) causes rapid growth of plasma oscillations which then become 
unstable. The same mechanism has been invoked in other astrophysical environments like particle acceleration in shocks \citep{1538-4357-543-1-L67,1538-4357-682-1-L5}.

By comparing the energy loss rates associated to plasma instabilities with the time scale of IC scattering, i.e.~the dominant mechanism of cascade development, 
the conclusion was drawn that the cascade might be suppressed on a short time scale which would imply that there is no need to require the existence of small 
intergalactic magnetic fields in order to explain the Fermi non-detection of GeV gamma-rays from Inverse Compton scattering. In addition, the energy deposited 
into the IGM could have dramatic consequences on its thermal history \citep{Chang:2011bf}.

In this paper we reanalyze the role of plasma instabilities adopting the formalism of \citet{0004-637X-758-2-102} and implementing these effects in a Monte 
Carlo simulation of the electromagnetic shower induced by TeV Blazars. Subsequently we are comparing our results with the predictions of the scenario in which 
only EGMF induced deflections are taken into account. In particular we focus on the TeV blazar 1ES 0229+200 located at $z=0.14$ since observational data are 
available at TeV \citep{2007A&A...475L...9A} as well as GeV \citep{Vovk:2011aa} energies. 

We structure our paper as follows: In Sec.~\ref{sec:PlasmEff} we introduce the formalism for taking into account plasma instabilities generated by electron 
propagation in the IGM. In Sec.~\ref{sec:SupprSpectr} we consider the impact of these effects on the modeling of TeV cascades from a specific source to 
determine the modification of the resulting photon spectrum and to assess whether these effects can account for the photon flux suppression at low energies. In 
addition we compare our results with models in which deflection by magnetic fields play a major role. Finally, our conclusions are drawn in Sec.~\ref{sec:Conclusions}.

\section{Plasma Effects on Electron/Positron Beams in the IGM} \label{sec:PlasmEff}
Following the formalism developed in \citet{0004-637X-758-2-102} here and later on, we model the electron/positron beam with number density $n_{\rm{beam}}$ 
resulting from the gamma-ray injection by the source and the IGM (consisting of protons and electrons with density~$n_{\rm{IGM}}$ as well as neutral atoms) as 
two streams of medium with different velocities. In such a setting it is possible for the so-called two-stream-like instabilities to grow on some specific time 
scale $\tau$. In this section we present a brief recap of the time scale derivations in both the linear and the non-linear regime.

Plasma instabilities are a well-known phenomenon which arises due to plasma oscillations \citep{0029-5515-1-1-002}: The time dependence of the propagation of 
electromagnetic waves is given by $\mathbf{E},\mathbf{B} \propto \exp\left( i \omega t \right)$. While for vacuum solutions of the Maxwell Equations the 
angular frequency $\omega$ is always real, propagation inside a medium may change the dispersion relation in such a way that also complex values for $\omega$ 
are possible. In particular, if $\mathfrak{Im}\left( \omega \right) < 0$ the expression $\exp\left( i \omega t \right)$ obtains an exponentially growing term 
which makes the setting unstable, therefore causing a plasma instability.

Considering the linear effects mentioned above, in the situation of interest here the linear oblique electrostatic instability has been found to be the 
dominant one. As we are dealing with a two-stream (i.e.~the relativistic electron beam and the IGM) setting, the dispersion relation is given by \citep{0004-637X-758-2-102,Godfrey1975}
\begin{equation} \label{DispRel}
1 = \frac{\omega_{p,e}^{2}}{\omega^{2}} + \frac{\omega_{\rm{beam}}^{2}}{\left( \omega - k_{\parallel} v \right)^{2}}
\end{equation}
where $\omega_{p,e}^{2} = 4 \pi e^{2} n_{\rm{IGM}}/m_{e}$ is the plasma frequency, $k_{\parallel}$ the wavenumber parallel to the propagation 
direction, $v$ the velocity of the beam electrons and positrons and 
\begin{equation}
\omega_{\rm{beam}}^{2} = \omega_{\parallel}^{2} \cos^{2}\theta + \omega_{\perp}^{2} \sin^{2}\theta
\end{equation}
an effective angular frequency depending on the longitudinal ($\omega_{\parallel}^{2} = 4 \pi e^{2} m_{e}^{2} n_{\rm{beam}} E_{e}^{-3}$) and transversal 
($\omega_{\perp}^{2} = 4 \pi e^{2} n_{\rm{beam}} E_{e}^{-1}$) angular frequency, respectively, as well as on the angle $\theta$ between the wave vector and the
direction of the flow, i.e.~$\cos\theta = k_{\parallel}/k$.

Eq.~\ref{DispRel} corresponds to the general form of the dispersion relation of a plasma carrying a current of electrons with velocity $v$, wave number $k$
as well as the ion and electron plasma angular frequencies $\omega_{p,i}$ and $\omega_{p,e}$, respectively. This general form is given by
\begin{equation}
1 = \frac{\omega_{p,i}^{2}}{\omega^{2}} + \frac{\omega_{p,e}^{2}}{\left( \omega - k v \right)^{2}}\,,
\end{equation}
for which a solution for $\omega = \omega_{\rm{max}}$, the angular frequency at maximum growth rate of the instability, is known to be \citep{0029-5515-1-1-002}
\begin{equation} \label{omegamax}
\begin{split}
\omega_{\rm{max}} &= \frac{k v}{1 + \alpha \exp\left(i \frac{\pi}{3} \right)} \\
&= \frac{k v}{\alpha^{2} + \alpha + 1} \left[ \left( \frac{\alpha}{2} + 1 \right) - i\frac{3^{\frac{1}{2}} \alpha}{2} \right]
\end{split}
\end{equation}
where $\alpha$ has to fulfill the conditions
\begin{equation}
\frac{2\alpha + 1}{\alpha^{3}\left( \alpha + 2 \right)} = \frac{\omega_{p,i}^{2}}{\omega_{p,e}^{2}},\,1 + \frac{3 \alpha^{2} + 2 \alpha + 1}{\alpha^{3}\left( \alpha + 2 \right)} = \frac{\left( k v \right)^{2}}{\omega_{p,e}^{2}}\,,
\end{equation}
i.e.~for Eq.~\ref{DispRel}
\begin{equation} \label{CondAlpha}
\frac{2\alpha + 1}{\alpha^{3}\left( \alpha + 2 \right)} = \frac{\omega_{p,e}^{2}}{\omega_{\rm{beam}}^{2}},\,1 + \frac{3 \alpha^{2} + 2 \alpha + 1}{\alpha^{3}\left( \alpha + 2 \right)} = \frac{\left( k_{\parallel} v \right)^{2}}{\omega_{\rm{beam}}^{2}}\,.
\end{equation}
Since we have $\omega_{p,e} \gg \omega_{\rm{beam}}$, it is $\omega_{p,e}^{2}/\omega_{\rm{beam}}^{2} \gg 1$ and therefore it follows from the first 
expression of Eq.~\ref{CondAlpha} that $\alpha$ has to be small such that via a Taylor Expansion Eq.~\ref{CondAlpha} may be rewritten as
\begin{equation} \label{CondAlpha1}
\frac{1}{2 \alpha^{3}} \simeq \frac{\omega_{p,e}^{2}}{\omega_{\rm{beam}}^{2}},\,\frac{1}{2 \alpha^{3}} \simeq \frac{\left( k_{\parallel} v \right)^{2}}{\omega_{\rm{beam}}^{2}}\,.
\end{equation}
which means that
\begin{equation} \label{kvomega}
k_{\parallel} v \simeq \omega_{p,e}\,.
\end{equation}
Performing a Taylor Expansion in $\alpha$ on Eq.~\ref{omegamax} and plugging into Eq.~\ref{kvomega} gives
\begin{equation} \label{omegamax1}
\begin{split}
&\mathfrak{Im}\left( \omega_{\rm{max}} \right) \simeq \frac{3^{\frac{1}{2}}}{2} \omega_{p,e} \alpha \stackrel{(\ref{CondAlpha1})}{\simeq} 3^{\frac{1}{2}} 2^{-\frac{4}{3}} \omega_{p,e}^{\frac{1}{3}} \omega_{\rm{beam}}^{\frac{2}{3}} \\
&= 3^{\frac{1}{2}} 2^{-\frac{4}{3}} \left(\frac{4 \pi e^{2} n_{\rm{IGM}}}{m_{e}}\right)^{\frac{1}{6}} \left( \omega_{\parallel}^{2} \cos^{2}\theta + \omega_{\perp}^{2} \sin^{2}\theta \right)^{\frac{1}{3}} \\
&= 3^{\frac{1}{2}} 2^{-\frac{4}{3}} \left(\frac{4 \pi e^{2} n_{\rm{IGM}}}{m_{e}}\right)^{\frac{1}{6}} \left[ \omega_{\perp}^{2} \left( 1 - v^{2} \cos^{2}\theta \right) \right]^{\frac{1}{3}} \\
&= \frac{3^{\frac{1}{2}} 2^{-\frac{4}{3}} \left( 4 \pi \right)^{\frac{1}{2}} e}{m_{e}^{\frac{1}{6}}} n_{\rm{IGM}}^{\frac{1}{6}} n_{\rm{beam}}^{\frac{1}{3}} E_{e}^{-\frac{1}{3}} \left( 1 - v^{2} \cos^{2}\theta \right)^{\frac{1}{3}}\,.
\end{split}
\end{equation}
Finally, taking into account the approximation in Eq.~\ref{kvomega} and considering ultrarelativistic electron/positron beams (i.e.~$v \simeq 1$), it can be 
shown that Eq.~\ref{omegamax1} becomes maximal for $\cos\theta = \left( 3/5 \right)^{\frac{1}{2}}$ \citep{0004-637X-758-2-102}.

Using this solution for the astrophysical scenario presented in this work, the electrostatic growth time is found to be
\begin{equation} 
\begin{split}
\tau_{\rm{e}} &\simeq 1.1 \times 10^{6}\,{\rm s}  \\
&\times \left( \frac{E_{e}}{10^{12}\,{\rm eV}} \right)^{\frac{1}{3}} \left( \frac{n_{\rm{beam}}}{10^{-22}\,{\rm cm^{-3}}} \right)^{-\frac{1}{3}} \left( \frac{n_{\rm{IGM}}}{10^{-7}\,{\rm cm^{-3}}} \right)^{-\frac{1}{6}}\,,
\end{split}
\end{equation}
where $E_{e}$ is the the energy of the electrons/positrons.

However, as mentioned above, we have to consider non-linear effects as well. This is done by calculating the total relaxation time $\tau_{\rm{r}}$ as 
\citep{Grognard1975,0004-637X-758-2-102}
\begin{equation} \label{RelElectro}
\tau_{\rm{r}} = 100 \tau_{\rm{e}} \xi^{-1}\,{\rm s}\,,
\end{equation}
introducing a dimensionless parameter $\xi \leq 1$ which has got a characteristic value for a particular non-linear effect and therefore accounts for its 
influence as discussed in the following.

The dominating effect in our setting is the modulation instability \citep{0004-637X-758-2-102}. It can be explained by the fact that in a turbulent medium ions 
scatter the oscillations caused by the beam such that they are transferred from the resonance to smaller wavenumbers. This, on the other hand, means that the 
energy is shifted to higher phase speeds \citep{GaleevJETP1977}. 

However, this effect occurs only if the beam density $n_{\rm{beam}}$ lies above a critical density $n_{\rm{crit}}$, namely
\begin{equation} 
\begin{split} \label{ncrit}
n_{\rm{beam}} > n_{\rm{crit}} &= 2.5 \times 10^{-25} {\rm cm^{-3}} \\
&\times \left( \frac{E_{e}}{10^{12}\,{\rm eV}} \right)^{-1} \left( \frac{n_{\rm{IGM}}}{10^{-7}\,{\rm cm^{-3}}} \right) \left( \frac{T}{10^{4}\,{\rm K}} \right)^{2}\,,
\end{split}
\end{equation}
where $T$ is the temperature of the IGM. The dissipation time scale here is given by
\begin{equation}
\begin{split}
\tau_{\rm{M}} &= 8.3 \times 10^{6}\,{\rm s} \left[ 1 + \frac{5}{4} \ln\left( \frac{T}{10^{4}\,{\rm K}} \right) - \frac{1}{4} \ln\left( \frac{n_{\rm{IGM}}}{10^{7}\,{\rm cm^{-3}}} \right)\right] \\
&\times \left( \frac{E_{e}}{10^{12}\,{\rm eV}} \right)^{\frac{1}{3}} \left( \frac{n_{\rm{beam}}}{10^{-22}\,{\rm cm^{-3}}} \right)^{-\frac{1}{3}} \left( \frac{n_{\rm{IGM}}}{10^{-7}\,{\rm cm^{-3}}} \right)^{-\frac{1}{6}}\,.
\end{split}
\end{equation}

If, however, we look at a weak blazar, i.e.~$n_{\rm{beam}}$ lies below $n_{\rm{crit}}$, then another, less efficient mechanism suppresses the cascade 
evolution, namely the non-linear Landau Damping (NLD). For this process the rate is known \citep{0029-5515-14-6-012}, so the factor $\xi$ in Eq.~\ref{RelElectro} 
can be explicitly calculated, resulting in
\begin{equation}
\begin{split}
\xi &= 2.1 \times 10^{-7} \times \left( \frac{E_{e}}{10^{12}\,{\rm eV}} \right)^{-\frac{4}{3}} \\
&\times \left( \frac{n_{\rm{beam}}}{10^{-22}\,{\rm cm^{-3}}} \right)^{-\frac{2}{3}} \left( \frac{n_{\rm{IGM}}}{10^{-7}\,{\rm cm^{-3}}} \right)^{\frac{2}{3}} \left( \frac{T}{10^{4}\,{\rm K}} \right)^{2}\,,
\end{split}
\end{equation}
which, plugged into Eq.~\ref{RelElectro}, gives
\begin{equation}
\begin{split} \label{tauNLD}
\tau_{\rm{NLD}} &\simeq 5.2 \times 10^{14}\,{\rm s} \left( \frac{E_{e}}{10^{12}\,{\rm eV}} \right)^{\frac{5}{3}} \\
&\left( \frac{n_{\rm{beam}}}{10^{-22}\,{\rm cm^{-3}}} \right)^{\frac{1}{3}} \left( \frac{n_{\rm{IGM}}}{10^{-7}\,{\rm cm^{-3}}} \right)^{-\frac{5}{6}} \left( \frac{T}{10^{4}\,{\rm K}} \right)^{-2} \,.
\end{split}
\end{equation}

In order to evaluate the impact of these damping effects on the development of an electromagnetic cascade, the corresponding time scales, 
$\tau_{\rm{M}}$ and $\tau_{\rm{NLD}}$, have to be compared to $\tau_{\rm{IC}}$, the Inverse Compton interaction time scale, which is given by
\begin{equation}
\tau_{\rm{IC}} = \frac{\lambda_{\rm{IC}}}{\beta_{e} c} \simeq 1.2 \times 10^{11} (1+z)^{-3} \, \rm{s}
\end{equation}
for the mean free path of Inverse Compton scattering $\lambda_{\rm{IC}}$, and to the IC cooling time
\begin{equation}
\tau_{{\rm IC,cool}} = \frac{E_{e}}{{\rm d}E_{e}/{\rm d}t} \simeq 3.87 \times 10^{13} \left( \frac{E_{e}}{10^{12}\,{\rm eV}} \right)^{-1} (1+z)^{-4} \, \rm{s}\,.
\end{equation}
while the average energy of the IC-produced photons for ultrarelativistic electrons considered here is given by
\begin{equation}
E_{\gamma} \simeq \frac{4}{3} E_{\rm{CMB}} \left( \frac{E_{e}}{m_{e}} \right)^{2}
\end{equation}
which for typical energies of CMB photons, $E_{\rm{CMB}} \simeq 7 \times 10^{-4}\,\rm{eV}$, gives
\begin{equation} \label{ICEgamma}
E_{\gamma} \simeq 4 \times 10^{9}\,\rm{eV}\,\left( \frac{E_{e}}{10^{12}\,\rm{eV}} \right)^{2}
\end{equation}

It should be noted here that these simple relations for the time scales are only valid in the Thomson regime, i.e.~for the case that the IC cross section can 
be assumed constant and equal to the Thomson cross section $\sigma_{{\rm T}} \simeq 6.55 \times 10^{-25}\,{\rm cm^{2}}$, which for CMB interactions is a valid 
approximation for electron/positron energies $E_{e} \leq 10^{13}\,{\rm eV}$.

To estimate a value for $n_{\rm{beam}}$, as it is directly connected to the luminosity and other intrinsic properties of the source which are not known to the 
full extent, we use an estimated upper limit given by \citet{0004-637X-752-1-22}:  
\begin{equation} \label{nBrod}
\begin{split}
n_{\rm{beam}} &\simeq 7.4 \times 10^{-22}\,{\rm cm^{-3}} \\
&\times \left( \frac{L}{10^{38}\,{\rm W}} \right) \left( \frac{E_{e}}{10^{12}\,{\rm eV}} \right) \left( \frac{1+z}{2} \right)^{3 \zeta - 4}
\end{split}
\end{equation}
where $L$ is the isotropic-equivalent luminosity and $\zeta = 4.5$ for $z<1$ is a parameter that can be inferred from the analysis of the local Universe star 
formation rate \citep{Kneiske2004}.
For the specific source used in the present work, 1ES 0229+200, we adopt $z=0.14$ and $L=10^{37.5}\,{\rm W}$~\citep{0004-637X-752-1-22}.

The time-scales comparison is shown in the left panel of Fig.~\ref{fig:timescales} for some typical parameter values. It can be seen that in the strong blazar 
regime, i.e.~$n_{\rm{beam}}>n_{\rm{crit}}$, the electromagnetic cascade is completely suppressed since $\tau_{\rm{r}}$ is several orders of magnitude smaller 
than $\tau_{\rm{IC}}$ which means that, due to the modulation instability, almost all electrons have been relaxed to a rather non-interactive state long way 
before they can produce a high energy photon by Inverse Compton interaction. However, for electron/positron beams fulfilling $n_{\rm{beam}}<n_{\rm{crit}}$ (the 
weak blazar condition) a cascade can still develop, even though it is partially suppressed by NLD.

Finally, some remarks should be made on the parameters used for the IGM. The IGM is not completely homogeneous (since it contains voids and overdensities), 
hence the values for $n_{\rm{IGM}}$ and $T$ are both not constant but undergo fluctuations. However, on average we can set $n_{\rm{IGM}} \simeq 10^{-7}\,{\rm cm^{-3}}$ 
and $T \simeq 10^{4}\,{\rm K}$ where for the latter, due to larger uncertainties, we adopt different values in the range of $T = 10^{3}\,{\rm K}...10^{5}\,{\rm K}$ 
as it has been shown to be the case for low-redshift IGM \citep{0004-637X-758-2-102,2013MNRAS.434.3293P}. 

After giving our definition for $n_{\rm{beam}}$ in Eq.~\ref{nBrod}, on the right hand side of Fig.~\ref{fig:timescales} we show how the transition energy 
between the weak and strong blazar regimes changes as function of the IGM temperature for sources with the same luminosity $L$ placed at different redshifts. 

It should be noted that the ideas and conclusions which have been presented in this section are in the focus of an ongoing debate. In particular in \citet{Miniati:2012ge} 
it has been argued that the kind of analysis used here is not applicable to electromagnetic cascades in the voids. By performing a kinetic treatment the 
authors claim to find that the instability growth is severely suppressed and therefore the relaxation time for the electrons/positrons inside the beam remains 
much larger than the IC interaction time. However, in a more recent paper \citep{Schlickeiser:2013eca} these objections have been addressed by refining the 
analysis and concluding that the results from \citet{0004-637X-758-2-102} are still valid. 

\begin{figure*}
\centering
\includegraphics[scale=0.58]{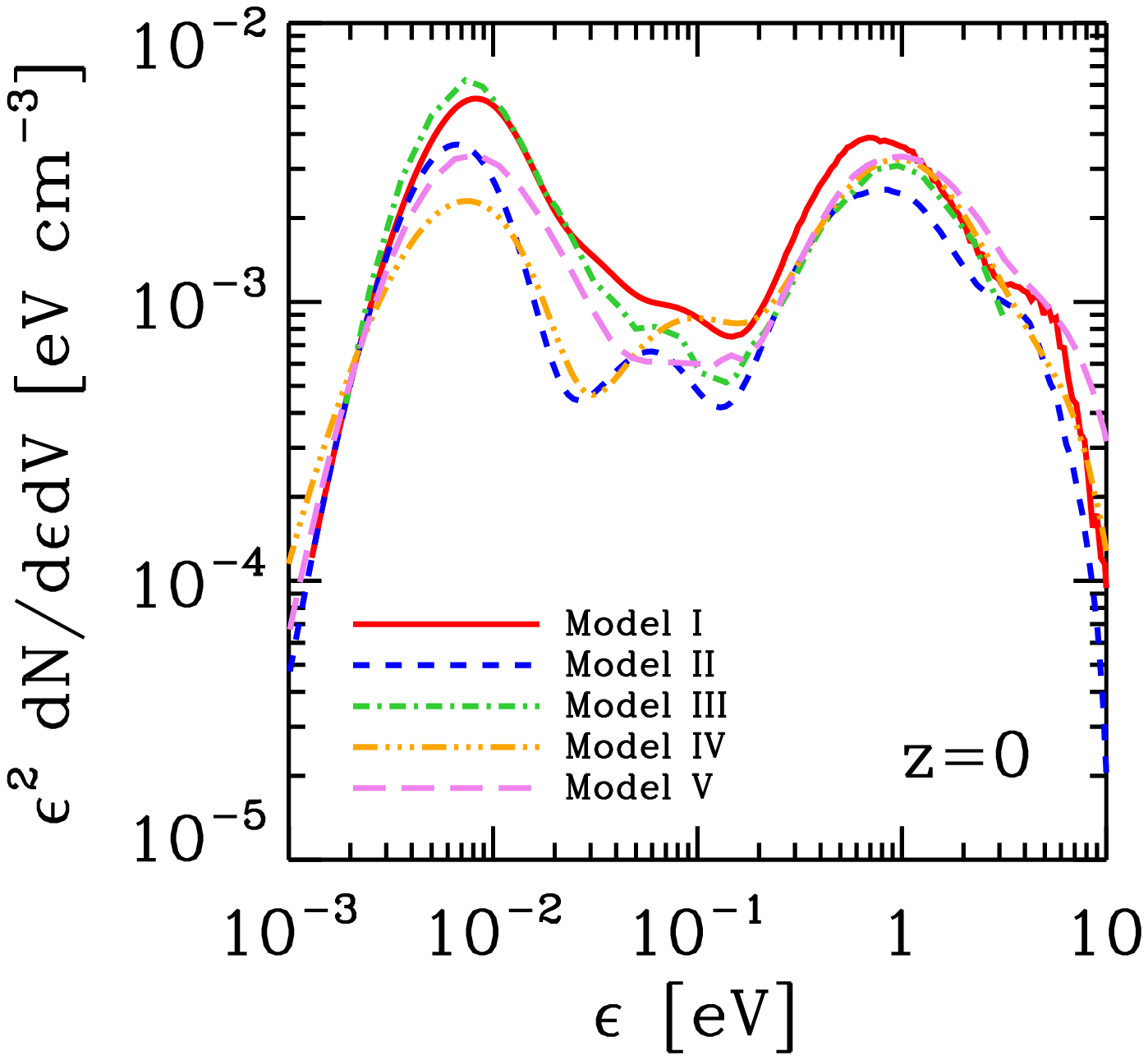}
\includegraphics[scale=0.58]{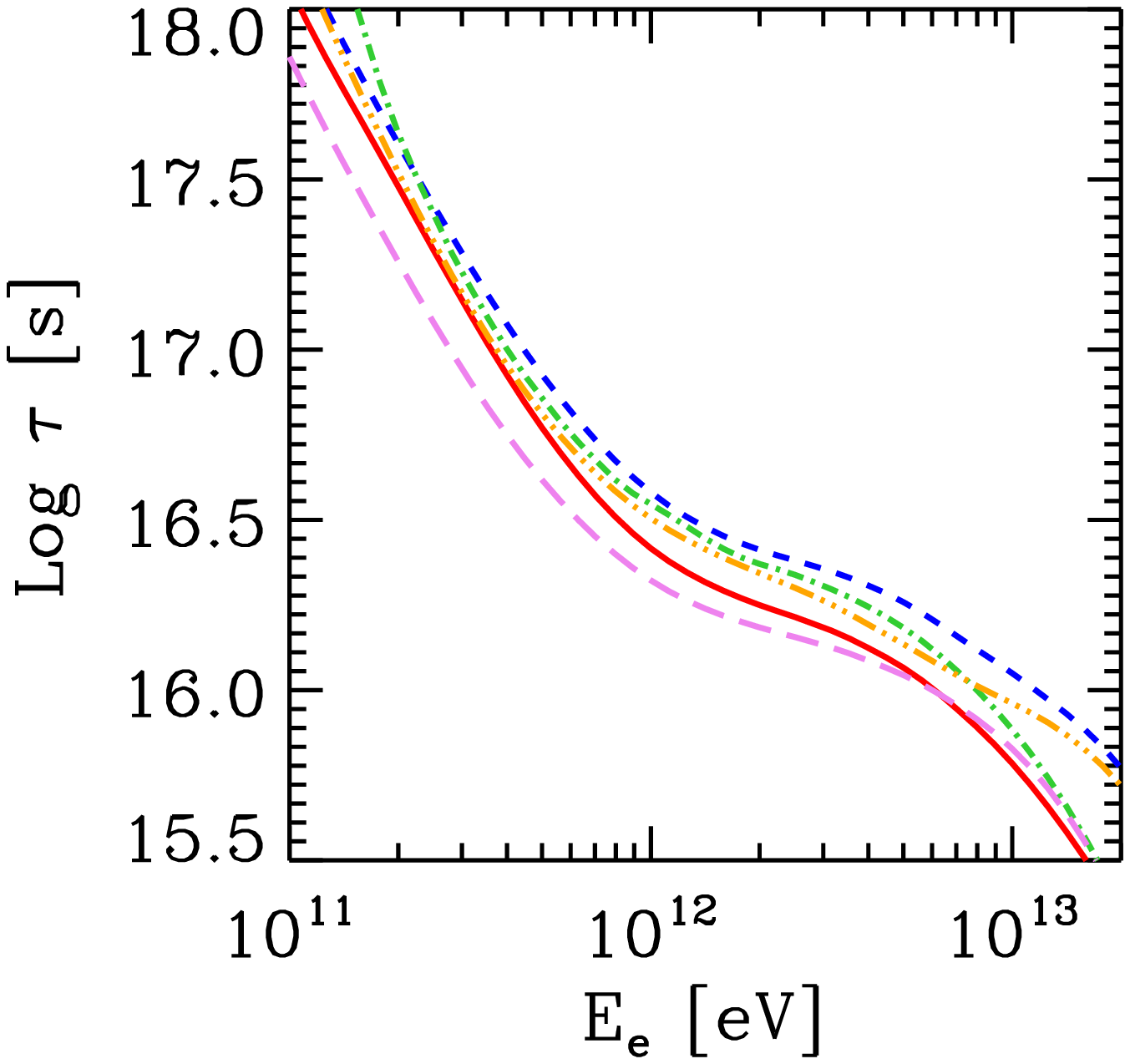}
\caption{
  \textit{Left panel}: The energy density of the infrared background at redshift $z=0$ for the various models (I-V) listed in Sec.~\ref{sec:SupprSpectr} 
  depending on the background photon energy $\epsilon$. \textit{Right panel}: The interaction time for pair production on the EBL for the same infrared 
  backgrounds.
} 
\label{fig:timescalesgamma}
\end{figure*}

\section{Suppression of Low Energy Photons in Observed Spectra of Blazars} \label{sec:SupprSpectr}
Using the considerations from the previous section, it is now possible to calculate the modification of the initial photon spectra from blazars induced by 
propagation in the IGM. To do so we make use of the ELMAG code \citep{2012CoPhC.183.1036K}, designed to simulate electromagnetic cascades via the Monte Carlo 
approach, including reactions with the photon background fields, synchrotron energy losses and deflections due to magnetic fields in the small angle 
approximation. In addition, we modified the code in order to take into account the interactions described in Sec.~\ref{sec:PlasmEff} during the cascade 
development.

To this end, each time an electron enters the cascade or undergoes an Inverse Compton scattering we calculate both the actual time until the next IC 
interaction and the relaxation time as described in the previous section. If the latter is the smaller one of the two, we assume that the electron does not 
contribute to the cascade anymore.  This can be done following the left panel of Fig.~\ref{fig:timescales}: There it is visible that at small energies the 
energy loss of electrons/positrons is dominated by plasma effects so that the contribution from Inverse Compton scattering of secondary pairs is very small. 

In order to obtain exhaustive results we compare the outcome for different infrared background models which can be found in literature: the best-fit model of 
\citet{Kneiske2004} (Model I), the lower-limit model of \citet{2010A&A...515A..19K} (Model II), the model presented in \citet{Franceschini:2008tp} (Model III), 
model C of \citet{0004-637X-712-1-238} (Model IV) and finally the semi-analytic model of \citet{Gilmore:2011ks} (Model V). The impact of the different 
approaches can be seen in Fig.~\ref{fig:timescalesgamma}.

The object we are investigating here is the BL Lac 1ES 0229+200. This source is of particular interest since for it both high energy data from HESS \citep{2007A&A...475L...9A} 
as well as a GeV data analysis from Fermi LAT \citep{Vovk:2011aa} are available. In particular, in \citet{Vovk:2011aa} the same object is used in order to 
constrain the EGMF and therefore a direct comparison with our results is possible.

\begin{figure}
\centering
\includegraphics[scale=0.563]{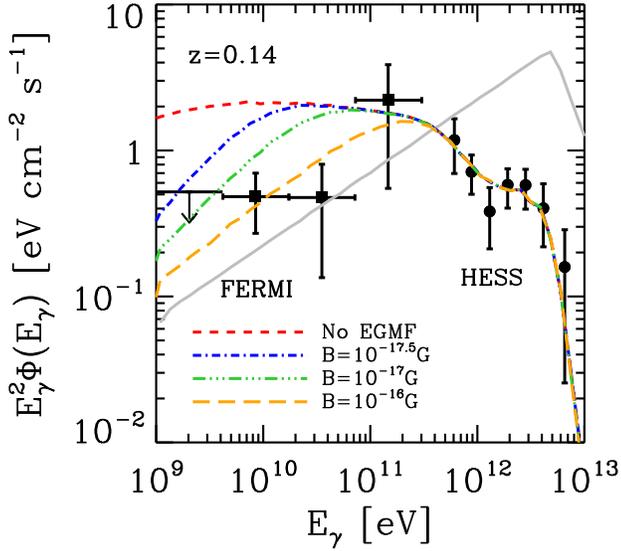}
\caption{
  Effect of EGMF on the photon spectrum neglecting the plasma effects of the IGM. Data points are shown for both HESS (circle) and Fermi LAT (square) while the 
  curves show the best fit of a Monte Carlo simulation for the HESS data. The chosen magnetic fields are $B=0\,{\rm G}$ (dashed, red), $B=10^{-17.5}\,{\rm G}$ 
  (dot-dashed, blue), $B=10^{-17}\,{\rm G}$ (dot-dot-dashed, green) and $B=10^{-16}\,{\rm G}$ (long-dashed, orange). In addition, the intrinsic spectrum is 
  given (solid, gray).
}
\label{fig:spectraMag}
\end{figure}

We first show in Fig.~\ref{fig:spectraMag} the result of our calculations for the case in which we neglect the effects caused by the IGM and assume a uniform 
magnetic field which affects the observed spectrum due to deflections of cascade leptons. Here, we assume an intrinsic spectrum with ${\rm d} N/{\rm d} E \propto E^{-\Gamma}$, $\Gamma = 1.5$ 
and an energy cutoff at $E_{\rm{cut}} = 5 \times 10^{12}\,{\rm eV}$. As shown also in~\citet{Vovk:2011aa}, this mechanism succeeds in explaining the low flux 
observed for photon energies in the GeV range.

\begin{figure*}
\centering
\includegraphics[scale=0.744]{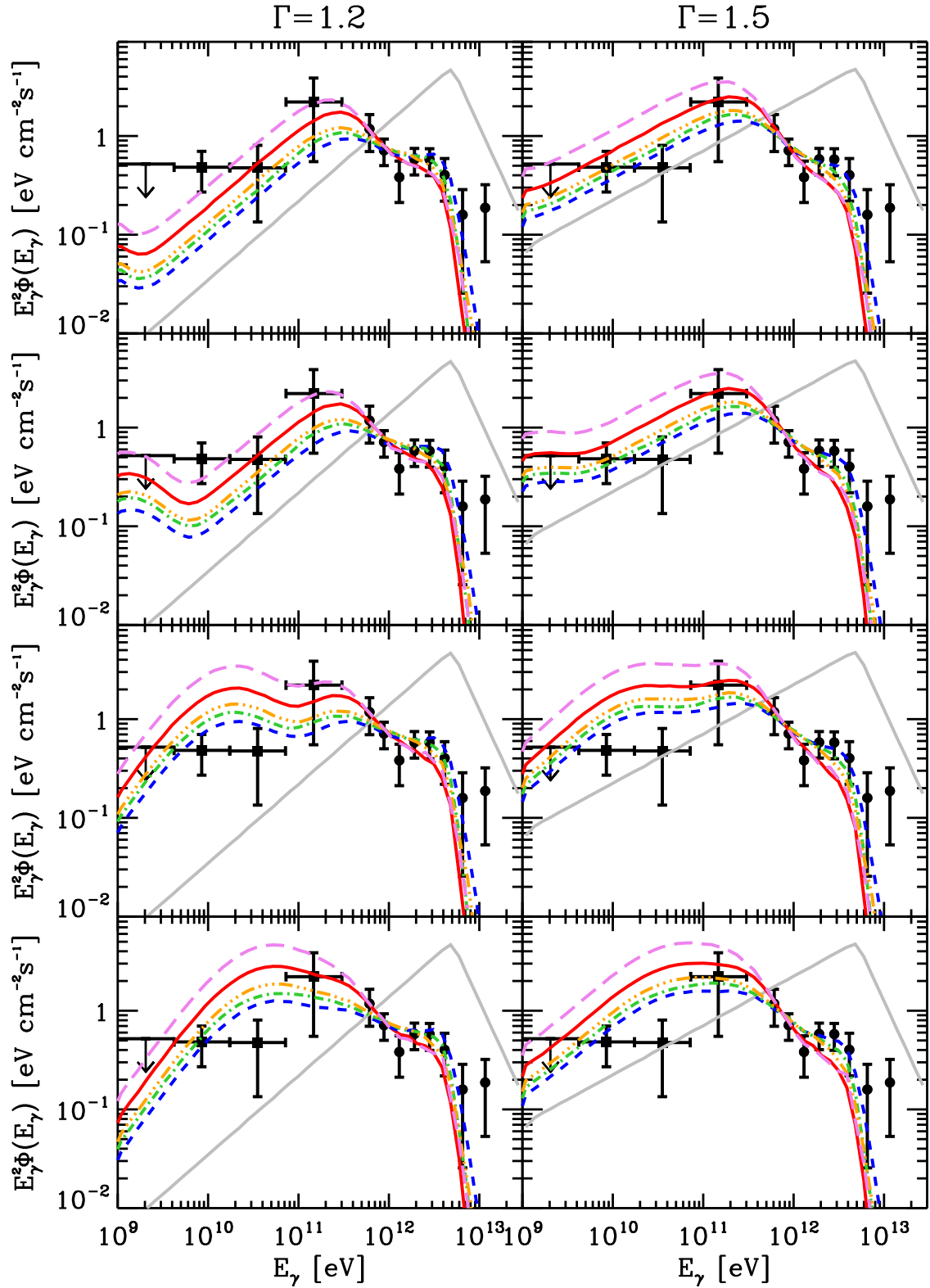}
\caption{
  Gamma-ray spectra of 1ES 0229+200 at $z=0.14$ for different temperatures and EBL models with the slope of the intrinsic spectrum (grey, solid) given by 
  $\Gamma=1.2$ (left) and~$\Gamma=1.5$ (right) in case IGM effects are taken into account and magnetic fields are supposed to be negligible. The temperatures 
  used here are $T=5 \times 10^{3}\,\rm{K}$, $T= 10^{4}\,\rm{K}$, $T=5 \times 10^{4}\,\rm{K}$, $T= 10^{5}\,\rm{K}$ (from top to bottom). Different line styles 
  stand for different EBL models as in Fig.~\ref{fig:timescalesgamma}.
}
\label{fig:spectraTemp}
\end{figure*}

However, as we present in Fig.~\ref{fig:spectraTemp}, even for a vanishing magnetic field the plasma effects due to the interaction with the IGM itself give a 
suppression of the flux in agreement with Fermi observations. It should be stressed here that this does not prove the non-existence of EGMF but rather weakens 
the role of GeV suppression for probing the lower limits for 
the magnetic field strength.

In Fig.~\ref{fig:spectraTemp} we put in evidence the effect of modifying the IGM temperature as it is the parameter giving the largest uncertainty. It can be 
seen that the dependence on $T$ has a non-trivial behavior as increasing it, according to Eq.~\ref{ncrit}, increases $n_{\rm{crit}}$ and therefore makes the 
\textit{weak} blazar regime dominating over the strong one and furthermore, once in this regime, a larger $T$ makes $\tau_{\rm{NLD}}$ become smaller, see 
Eq.~\ref{tauNLD}, and therefore increases the impact of NLD. 

This fact is reflected in Fig.~\ref{fig:spectraTemp}: For low temperatures the spectrum at small energies basically reproduces the slope of the intrinsic 
spectrum since the system operates in the strong blazar regime and therefore only the primary photons emitted by the source, i.e.~photons which have not 
created pairs during their propagation, reach the observer while the energy of the produced positron/electron pairs is dissipated away before they can perform 
an IC scattering. 

For higher temperatures an additional peak in the spectrum appears due to the influence of the weak blazar regime. The energy at which this peak is situated 
increases with temperature such that, for $T=10^{5}\,\rm{K}$, it dominates the spectrum which can be especially seen by the fact that the slope does not 
correspond to the intrinsic one anymore. This spectral feature can be explained by the fact that, as can be seen from Fig.~\ref{fig:timescales}, the energy at 
which the transition between the two regimes takes place, $E_{\rm{crit}}$, is the higher the higher the temperature is. All electrons with a energy lower than 
that therefore are governed by the weak blazar regime such that IC scattering is rather effective. 

For example, taking $T=5 \times 10^{4}\,\rm{eV}$, in the condition of 1ES 0229+200 we have $E_{\rm{crit}} \simeq 2 \times 10^{12}\,\rm{eV}$ and therefore, 
according to Eq.~\ref{ICEgamma}, the average energy of photons produced via IC is given by $E_{\gamma} \simeq 2 \times 10^{10}\,\rm{eV}$ which is approximately 
the energy of the additional peak as confirmed in Fig.~\ref{fig:spectraTemp} in the panel corresponding to $T=5 \times 10^{4}\,\rm{eV}$. In conclusion, the 
peak as well as the hardening of the spectrum for small energies may be explained by IC up-scattered photons.

However, in order to reproduce the HESS and Fermi LAT data, our model prefers low IGM temperature which, in turn, would mean that the slope of the observed 
spectrum would be much closer to the intrinsic one. For the intrinsic slope, two cases are presented here, $\Gamma=1.2$ and $\Gamma=1.5$ (see Fig.~\ref{fig:spectraTemp}).
We show that both values allow a good data fit, although giving different constraints on the EBL model. It might serve as another test of the different models 
in the future. As a consistency check we also used an extreme case of $\Gamma=1.8$ which gives a rather large flux at low energies though. In fact, the values 
presented here are only representative and a more dedicated analysis is required to estimate the allowed range for the initial spectrum.

\begin{figure}
\centering
\includegraphics[scale=0.563]{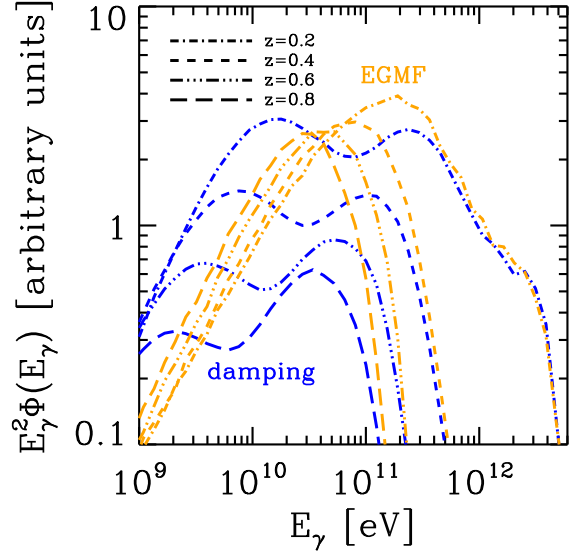}
\caption{
  The dependence of the observed blazar spectrum on the source redshift. The intrinsic spectrum has been chosen to have $\Gamma=1.2$. The blue lines give the 
  case including only the plasma effects for $T=5\times 10^{4}$, orange lines show the case in which particles are deflected by a magnetic field with $B=10^{-16}\,\rm{G}$ 
  neglecting the role of IGM.
}
\label{fig:spectraz}
\end{figure}

Finally, it is important to have a look at the dependence of the observed photon spectrum on the source redshift. In Fig.~\ref{fig:spectraz} this is done, on 
the one hand, by considering the role of IGM induced plasma instabilities and, on the other hand, neglecting these effects, by applying the deflections by a 
magnetic field. It should be noted here that we have used the same values for the IGM and for the magnetic field which can successfully be used to reproduce 
the observations in the specific case of 1ES 0229+200 with the EBL Model II (cf.~Figs.~\ref{fig:spectraMag} and \ref{fig:spectraTemp}). 

As one can see both give a flux suppression at low energies even for far away sources and also the energy cutoff remains almost the same. However, if plasma 
effects are operating, the observed flux at low energies, even for distant blazars, is the higher the lower the redshift of the source is, while for the 
deflection by magnetic fields it is the other way around. These trends are due to the different behaviors of the high energy (i.e.~$\simeq$ TeV) electrons: 
While the deflection by magnetic fields does basically not affect their propagation due to their large Larmor radii, the induced instabilities are particularly 
strong in that energy range (cf.~Fig.~\ref{fig:timescales}). Therefore, while in the former scenario the high energy electrons \emph{do} contribute to the 
cascade (for longer distances even repeatedly), and thus the low energy flux increases with the redshift, in the case of plasma instabilities their propagation 
is almost completely suppressed.
 
\section{Conclusions}
\label{sec:Conclusions}
While propagating in the IGM, TeV gamma-rays from distant blazars produce relativistic electron-positron beams by interacting with the EBL. The generated pair 
beam distribution is unstable due to two-stream instabilities in the unmagnetized IGM. Based on the formalism developed in \citet{0004-637X-752-1-22,0004-637X-758-2-102} 
and \citet{Schlickeiser:2013eca} we have properly modeled the ultrarelativistic pair beam produced in the IGM by multi-TeV gamma-ray photons from blazars. To 
do so the physical properties of the pair beam are determined through a Monte Carlo model of the electromagnetic cascade.

In summary we confirm that instabilities due to interactions of relativistic electron/positron beams with the IGM can be effective, so that the generation of 
Inverse Compton scattered photons by the pair beam is significantly suppressed. By comparing the spectrum resulting from this model to the one predicted in the 
scenario in which deflections due to EGMF are invoked to explain the observations, we show that both are compatible with existing data, the former for 
$B \gtrsim 10^{-16}\,\rm{G}$ and the latter for $T \lesssim 5 \times 10^{4}\,\rm{K}$. Therefore there is no need to require the existence of small 
intergalactic magnetic fields to explain the observed suppression in the GeV range of blazar spectra.
In case of plasma instabilities a second peak in the gamma-ray spectrum is predicted due to the transition between the weak and the strong regime. As this is a 
rather distinct feature, the analysis of (future) source observations at different redshifts $z\lesssim 1$ may be used to distinguish between the two energy 
loss scenarios. Upcoming observations by HESS-II \citep{Becherini:2012yq} and CTA \citep{Reimer:2012vw} would therefore help to shed light on the complex 
interaction of TeV gamma-rays with the IGM.  

Finally, implementing a detailed study of the IGM properties (as we aim to do in a following work) in order to determine the effect of environment 
inhomogeneities will further improve our understanding of the role of IGM in modeling blazar observations.

\section*{Acknowledgments}
We would like to thank Michael Kachelrie{\ss}, Steffen Krakau and Reinhard Schlickeiser for fruitful discussions and valuable comments. This work was supported 
by the Deutsche Forschungsgemeinschaft through the collaborative research centre SFB 676, by the Helmholtz Alliance for Astroparticle Phyics (HAP) funded by 
the Initiative and Networking Fund of the Helmholtz Association, and by the State of Hamburg through the Collaborative Research program ``Connecting Particles 
with the Cosmos''.

\label{lastpage}
\end{document}